# Combined Piezoelectrooptic Effect in Rochelle Salt at the Phase Transition

Vlokh R., Mys O. and Kostyrko M.

Institute of Physical Optics, 23 Dragomanov Str., 79005 L'viv, Ukraine

**Abstract**

The results of study of piezoelectrooptic (PEO) effect in the course of ferroelectric-ferroelastic phase transition in Rochelle salt crystals are presented. The coefficient of the combined effect is obtained from measurements of the changes in the electrooptic coefficients under the action of mechanical stress and the changes in the piezooptic coefficients under the action of electric field (or spontaneous polarization). It is shown experimentally that the values of both coefficients are the same, as predicted by the theory. The temperature dependence of the coefficient of combined PEO effect is obtained. Its anomalous behaviour at the phase transition appears due to the dielectric permitivity anomaly. It is also demonstrated that the change in the piezooptic coefficients at the phase transition in Rochelle salt crystals may be satisfactorily explained as a result of linear and quadratic PEO effect induced by spontaneous polarization.

## Introduction

The combined piezoelectrooptic (PEO) effect consists in changes of the optical-frequency dielectric impermeability constant $\Delta B_{ij}$ (or the refractive indices $B_{ij} = \left(\frac{1}{n^2}\right)_{ij}$) of crystals under the mutual influence of electric field $E_m$ and mechanical stress $\sigma_{kl}$ [1]. It is described with the relation:

$$\Delta B_{ij} = N_{ijklm}\sigma_{kl}E_m + M_{ijklmn}\sigma_{kl}E_m E_n, \qquad (1)$$

where $N_{ijklm}$ is a fifth-rank polar tensor with the internal symmetry $[V^2]^2 V$, while $M_{ijklmn}$ is a polar tensor of a rank six with the internal symmetry $[V^2]^3$. It follows from the symmetry of the $N_{ijklm}$ tensor that the PEO effect linear in electric field can exist only in non-centrosymmetric crystals. This effect may be presented also as a change in the electrooptic (EO) coefficients under the action of mechanical stress:

$$\Delta r_{ijm} = N_{ijmkl}\sigma_{kl}, \qquad (2)$$

or, equivalently, a change in the piezooptic (PO) coefficients under the action of electric field:

$$\Delta \pi_{ijkl} = N_{ijklm}E_m. \qquad (3)$$

The PEO effect is not the only combined effect of the parametric crystal optics. One can remind a piezoelectrogyration effect found by *H-J.Weber et al* [2] and a so-called "magnetoelectrogyration" effect studied by *W.Kaminsky et al* [3] in the alum crystals. Furthermore, the changes in the Faraday coefficients under the electric field and spontaneous polarization in $Pb_5Ge_3O_{11}$ crystals observed by *O.Vlokh et al* [4], which have been later explained in [5] and described also by *I.Chupis* [6], seem to be a relevant example, too. Besides, the effects just mentioned describe the changes in imaginary part of the dielectric permeability tensor induced by mutual influence of different fields, while the effect of PEO (i.e., the change in the refractive indices) is related to real part of this tensor and so should be greater.

One of particular manifestations of the PEO effect in the course of phase transitions in the proper ferroelectric crystals may be a change in the PO coefficients occurring with the onset of spontaneous polarization $P^{(s)}_m$ [7]:

$$\Delta \pi_{ijkl} = N^{(s)}_{ijklm} P^{(s)}_m, \qquad (4)$$

where $N^{(s)}_{ijklm}$ is taken for the paraelectric phase. The same should be true of the EO coefficients in the proper ferroelastic crystals. The change in the EO and PO coefficients at the phase transition in ferroelectric-ferroelastic crystals should depend on both the spontaneous polarization and deformation (as far as we know, the combined PEO effect has been studied only in proper ferroelectrics up to now). On the other side, due to piezoelectric coupling between the spontaneous polarization and spontaneous deformation in the ferroelectric-ferroelastic crystals, one can expect appearance of links between the coefficients of the combined PEO effect and the usual Pockels or Kerr coefficients and the PO coefficients (see, e.g., [8] where the relations between the PO, EO, piezogyration and electrogyration effects have been obtained and experimentally verified for ferroelectrics-ferroelastics). Thus, the present work is devoted to study of PEO effect at ferroelectric-ferroelastic phase transition on the example of Rochelle salt crystals.

**Experimental**

Rochelle salt crystals ($NaKC_4H_4O_6 \times 4H_2NH_2O$) possess two ferroelectric-ferroelastic phase transitions with the change of point symmetry 222F2F222 at $T_{c1}$=297K and $T_{c2}$=255K (see, e.g., [9]). The first transition is a direct phase transformation to the ferroelectric-ferroelastic phase with the point symmetry group 2, while the second one is re-entrant and results in repairing the paraphase with the point symmetry 222.

$NaKC_4H_4O_6 \times 4H_2NH_2O$ single crystals with the flat plane-parallel faces perpendicular to <100> (the axis of spontaneous polarization), <011> and <01$\bar{1}$> directions were polished with the diamond powders and paste. The radiation of He-Ne laser with the wavelength of 632.8nm was propagated along <01$\bar{1}$> direction. The electric field was directed along <100> axis using the copper electrodes and the mechanical stress was applied along <011> direction. The crystal was placed into a specially designed furnace, providing the accuracy of temperature stabilization of about 0.1K. The changes in the birefringence under the application of electric field were measured with the Senarmont method for different magnitudes of applied mechanical stresses, and vice versa. We studied the behaviour of the PEO effect only at the high-temperature phase transition, i.e. the direct transformation to the ferroelectric-ferroelastic phase. The PO and the EO coefficients were calculated for the case of single-domain state of samples, i.e. for the data taken from outside the hysteresis loop region.

**Results and discussion**

Let us analyse the changes in the optical indicatrix for the point groups 222 and 2 and the mentioned experimental geometry. The PO and EO tensors for these groups are available, e.g., in [10], while the fifth- and sixth-rank tensors are still necessary to derive. The columns of these tensors that will be used further on have the following form:

a) for the point group 222

|  | $\sigma_2 E_1$ | $\sigma_3 E_1$ | $\sigma_4 E_1$ |
|---|---|---|---|
| $\Delta B_1$ | 0 | 0 | $N_{141}$ |
| $\Delta B_2$ | 0 | 0 | $N_{241}$ |
| $\Delta B_3$ | 0 | 0 | $N_{341}$ , |
| $\Delta B_4$ | $N_{421}$ | $N_{431}$ | 0 |
| $\Delta B_5$ | 0 | 0 | 0 |
| $\Delta B_6$ | 0 | 0 | 0 |

|        | $\sigma_2 E_1 E_1$ | $\sigma_3 E_1 E_1$ | $\sigma_4 E_1 E_1$ |
|--------|---------|---------|---------|
| $\Delta B_1$ | $M_{121}$ | $M_{131}$ | 0 |
| $\Delta B_2$ | $M_{221}$ | $M_{231}$ | 0 |
| $\Delta B_3$ | $M_{321}$ | $M_{331}$ | 0 |
| $\Delta B_4$ | 0 | 0 | $M_{441}$ |
| $\Delta B_5$ | 0 | 0 | 0 |
| $\Delta B_6$ | 0 | 0 | 0 |

;

|        | $\sigma_2 E_1 E_1$ | $\sigma_3 E_1 E_1$ | $\sigma_4 E_1 E_1$ |
|--------|---------|---------|---------|
| $\Delta B_1$ | $M_{121}$ | $M_{131}$ | $M_{141}$ |
| $\Delta B_2$ | $M_{221}$ | $M_{231}$ | $M_{241}$ |
| $\Delta B_3$ | $M_{321}$ | $M_{331}$ | $M_{341}$ |
| $\Delta B_4$ | $M_{421}$ | $M_{431}$ | $M_{441}$ |
| $\Delta B_5$ | 0 | 0 | 0 |
| $\Delta B_6$ | 0 | 0 | 0 |

.

b) for the point group 2 (the two-fold symmetry axis being parallel to $x$)

|        | $\sigma_2 E_1$ | $\sigma_3 E_1$ | $\sigma_4 E_1$ |
|--------|---------|---------|---------|
| $\Delta B_1$ | $N_{121}$ | $N_{131}$ | $N_{141}$ |
| $\Delta B_2$ | $N_{221}$ | $N_{231}$ | $N_{241}$ |
| $\Delta B_3$ | $N_{321}$ | $N_{331}$ | $N_{341}$ |
| $\Delta B_4$ | $N_{421}$ | $N_{431}$ | $N_{441}$ |
| $\Delta B_5$ | 0 | 0 | 0 |
| $\Delta B_6$ | 0 | 0 | 0 |

,

Under the mutual influence of the electric field $E_1$ and the mechanical stress $\sigma_{<011>}$ (such the stress is coupled with the stress tensor components as $\sigma_3 = \sigma_2 = \frac{1}{2}\sigma_{<011>}$ and $\sigma_4 = -\frac{1}{2}\sigma_{<011>}$) in the paraphase of $NaKC_4H_4O_6 \times 4H_2NH_2O$, the optical indicatrix equation may be written as

$$(B_1 + \pi_{12}\sigma_2 + \pi_{13}\sigma_3 + N_{141}\sigma_4 E_1 + M_{121}\sigma_2 E_1^2 + M_{131}\sigma_3 E_1^2)x^2 +$$
$$(B_2 + \pi_{22}\sigma_2 + \pi_{23}\sigma_3 + N_{241}\sigma_4 E_1 + M_{221}\sigma_2 E_1^2 + M_{231}\sigma_3 E_1^2)y^2 +$$
$$(B_3 + \pi_{32}\sigma_2 + \pi_{33}\sigma_3 + N_{341}\sigma_4 E_1 + M_{321}\sigma_2 E_1^2 + M_{331}\sigma_3 E_1^2)z^2 +$$
$$2r_{41}E_1 zy + 2\pi_{44}\sigma_4 zy + 2N_{421}\sigma_2 E_1 zy + 2N_{431}\sigma_3 E_1 zy + 2M_{441}\sigma_4 E_1^2 zy = 1 \quad (5)$$

After considering separately all the effects that are included in Eq. (5), one can write out the particular equations for the increments of optical birefringence due to the EO effect:

$$\delta(\Delta n)_{23} = \sqrt{2}\left(\frac{n_3 n_2}{\sqrt{n_3^2 + n_2^2}}\right)^3 r_{41} E_1, \quad (6)$$

due to the PO effect:

$$\delta(\Delta n)_{23} = \sqrt{2}\left(\frac{n_3 n_2}{\sqrt{n_3^2 + n_2^2}}\right)^3 \pi_{44}\sigma_4 =$$
$$= -\frac{\sqrt{2}}{2}\left(\frac{n_3 n_2}{\sqrt{n_3^2 + n_2^2}}\right)^3 \pi_{44}\sigma_{<011>} \quad (7)$$

and, finally, due to the PEO effect:

$$\delta(\Delta n)_{23} = \sqrt{2}\left(\frac{n_3 n_2}{\sqrt{n_3^2 + n_2^2}}\right)^3 \times$$
$$\times (N_{421}\sigma_2 E_1 + N_{431}\sigma_3 E_1 + M_{441}\sigma_4 E_1^2) \quad (8)$$

or

$$\delta(\Delta n)_{23} = \frac{\sqrt{2}}{2}\left(\frac{n_3 n_2}{\sqrt{n_3^2 + n_2^2}}\right)^3 \times$$
$$\times \left((N_{421} + N_{431})\sigma_{<011>}E_1 - M_{441}\sigma_{<011>}E_1^2\right) \quad (8a)$$

where $n_2$ and $n_3$ are the principal refractive indices. In Eqs. (7) and (8a), we neglect the change in the principal refractive indices due to the electric field and mechanical stress, while their influence on the calculation of $\delta(\Delta n)_{32}$ is very small. We also neglect their temperature variation and use the mean value of the refractive indices $n \approx 1.49$. Eq. (6) is also correct for the ferroelectric-ferroelastic phase, whereas Eqs. (7) and (8a) should be rewritten to the forms

$$\delta(\Delta n)_{23} = \sqrt{2}\left(\frac{n_3 n_2}{\sqrt{n_3^2 + n_2^2}}\right)^3 \times$$
$$\times \left(\pi_{42} + \pi_{43} - \pi_{44}\right)\sigma_{<011>} \quad (9)$$

and

$$\delta(\Delta n)_{23} = \sqrt{2}\left(\frac{n_3 n_2}{\sqrt{n_3^2+n_2^2}}\right)^3 \times$$
$$\times (N_{421}\sigma_2 E_1 + N_{431}\sigma_3 E_1 + N_{441}\sigma_4 E_1) =$$
$$= \frac{\sqrt{2}}{2}\left(\frac{n_3 n_2}{\sqrt{n_3^2+n_2^2}}\right)^3 \times \qquad (10)$$
$$\times ((N_{421}+N_{431})\sigma_{<011>}E_1 - N_{441}\sigma_{<011>}E_1)$$

Notice that we neglect in Eq. (10) the combined of the effect quadratic in the electric field. Taking into account Eqs. (2) and (3), we may write the changes in the EO coefficient $r_{41}$ dependent on mechanical stress for the both structural phases, respectively, for the paraphase:

$$\Delta r_{41} = N_{421}\sigma_2 + N_{431}\sigma_3 =$$
$$= \frac{1}{2}(N_{421}+N_{431})\sigma_{<011>}, \qquad (11)$$
$$N_{eff} = (N_{421}+N_{431})$$

and the ferroelectric-ferroelastic phase:

$$\Delta r_{41} = N_{421}\sigma_2 + N_{431}\sigma_3 + N_{441}\sigma_4 =$$
$$= \frac{1}{2}(N_{421}+N_{431}-N_{441})\sigma_{<011>} \qquad (12)$$
$$N_{eff} = (N_{421}+N_{431}-N_{441})$$

The corresponding changes in the PO coefficients dependent on the electric field are

$$\Delta \pi_{eff} = (\Delta\pi_{42}+\Delta\pi_{43}) =$$
$$= \frac{1}{2}(N_{421}+N_{431})E_1 \qquad (13)$$

for the paraphase and

$$\Delta \pi_{eff} = (\Delta\pi_{42}+\Delta\pi_{43}-\Delta\pi_{44}) =$$
$$= \frac{1}{2}(N_{421}+N_{431}-N_{441})E_1 \qquad (14)$$

for the ferroelectric-ferroelastic phase. As seen from Eqs. (11) and (13), the increment of the EO coefficient ($\Delta r_{41}$) and that of the PO coefficient ($\Delta\pi_{44}$) at the phase transition in Rochelle salt crystals do not depend on the spontaneous deformation $e_4$ or the spontaneous polarization $P^s_1$, because the $N^{(s)}_{441}$ value is equal to zero for the point group 222. The coefficient $\pi_{44}$ can manifest a quadratic dependence on $P^{(s)}_1$. Thus, the temperature dependence of the increment of effective PO coefficient, which is measured in our experimental geometry, may be described with the relation

$$\Delta \pi_{eff} = (\pi_{42}+\pi_{43}-\pi_{44}) =$$
$$= \frac{1}{2}(N^{(s)}_{421}P^{(s)}_1 + N^{(s)}_{431}P^{(s)}_1 - M^{(s)}_{441}P^{2(s)}_1), \qquad (15)$$

where $N^{(s)}_{431}$, $N^{(s)}_{421}$ and $M^{(s)}_{441}$ are the coefficients for the paraphase. The temperature dependence of the effective PO coefficient measured on electrically shorted sample and calculated on the basis of Eqs. (7) and (9) is presented in Figure 1.

One can see that the increment of this coefficient appearing at the phase transition depends linearly on temperature (if we do not take into account a diffuse character of the phase transition). On the other hand, the dependence of $\pi_{eff}$ on the spontaneous polarization (the data for temperature dependence of the spontaneous polarization are taken from [11]) is nonlinear (see Figure 2a).

It is possible to distinguish between the linear and quadratic parts of this dependence (see Figure 2b) and calculate the corresponding coefficients of the PEO effect induced by spontaneous polarization, using Eq. (15). It is found that the coefficients are equal to $(N^{(s)}_{421}+N^{(s)}_{431})=1.25\times10^{-9} m^4/NC$ and $M^{(s)}_{441}=9.4\times10^{-7} m^6/NC^2$ for the paraphase. The linear part of the dependence is connected with the appearance of coefficients $\pi_{42}$ and $\pi_{43}$ in the ferroelectric-ferroelastic phase.

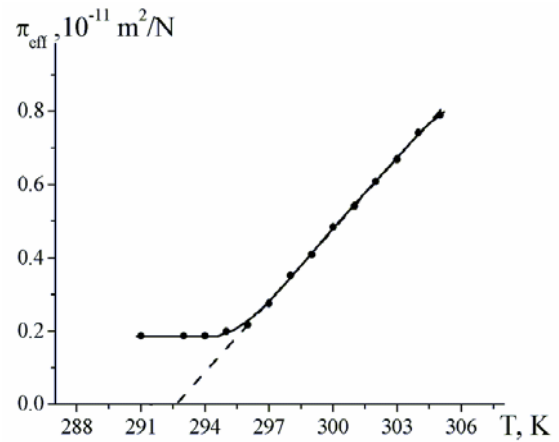

**Fig. 1.** Temperature dependence of the effective PO coefficient for NaKC$_4$H$_4$O$_6$×4H$_2$NH$_2$O crystals ($\lambda$=632.8nm).

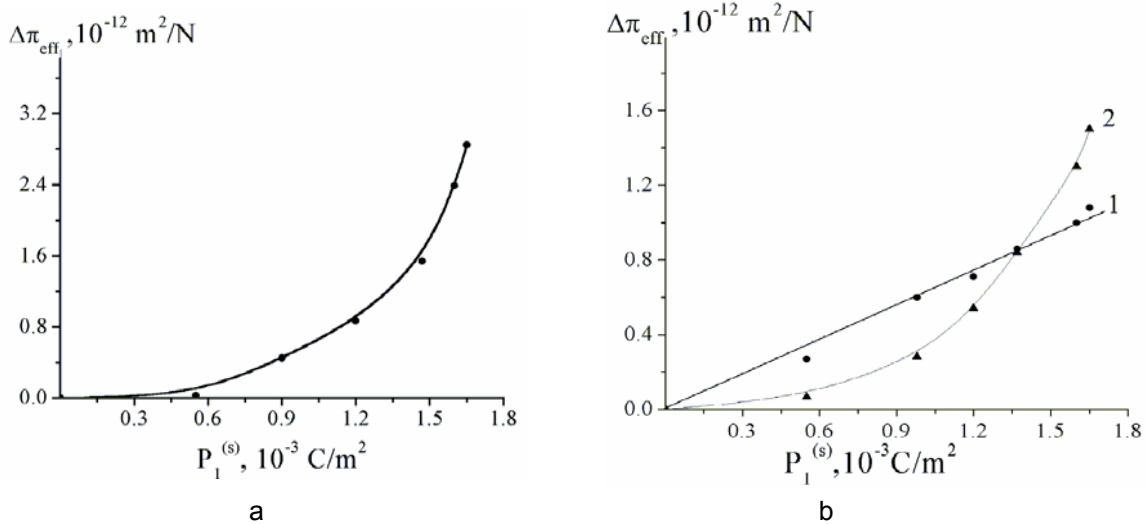

**Fig. 2.** Dependence of the effective PO coefficient on the spontaneous polarization for $NaKC_4H_4O_6 \times 4H_2NH_2O$ crystals: (a) according to the experimental data and (b) after separating the linear (1) and quadratic (2) PEO effects ($\lambda$=632.8nm).

The PO hysteresis loops obtained at 295K at different values of the biasing field are presented in Figure 3. With increasing biasing field, one can observe the increase in the size of hysteresis, followed by its subsequent decrease. It is necessary to note that an electromechanical hysteresis should also exist in ferroelectric-ferroelastic crystals. Therefore, the observed effect becomes quite understandable i.e., the application of electric field leads to changing the spontaneous deformation in the limit of "butterfly-like" hysteresis. On the other side, the dependence of changes in the PO coefficient on the electric field may be obtained from the slopes of linear dependences of the birefringence induced by mechanical stresses at different electric field, out of the hysteresis loop region. This dependence, which is represented in Figure 4, turns out to be linear.

Using the data of Figure 4 and Eq. (14), it is possible to determine the effective coefficient of the PEO effect. Its value equals to $N_{eff}=1.55\times10^{-17}$ Vm/N.

The temperature dependences of EO

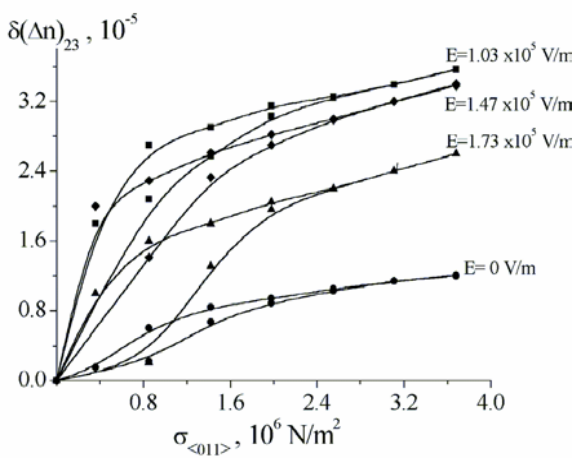

**Fig. 3.** Piezooptic hysteresis loops obtained at different values of biasing field for the Rochelle salt crystals (T=295K, $\lambda$=632.8nm).

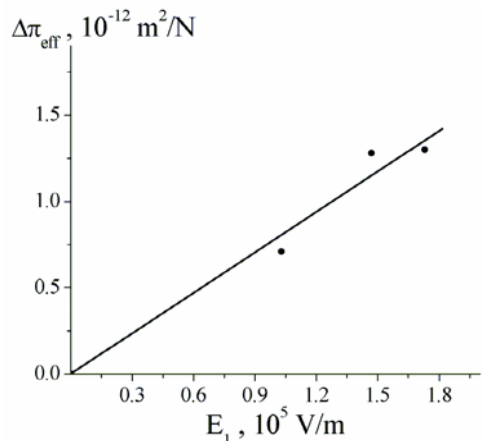

**Fig. 4.** Dependence of the effective PO coefficient on the electric field for $NaKC_4H_4O_6 \times 4H_2NH_2O$ crystals at 295K ($\lambda$=632.8nm).

coefficient in the vicinity of phase transition are presented in Figure 5 for different mechanical stresses. As one can easily see, the magnitude of EO coefficient that has been calculated with the aid of Eq. (6) increases, due to the PEO effect, when the compressive stress is applied. The effective coefficient of the combined effect is $N_{eff}=1.46\times10^{-17}$Vm/N at 295K (see Eqs. (11) and (12)). It is worthwhile that we have got rather close values for the coefficients $N_{eff}$ determined from the measurements of both the PO coefficient change under the electric field and the EO coefficient change under the mechanical stress.

It is seen from Figure 5 that the difference between the curves of EO coefficients obtained at different stresses gradually disappears approximately at 305K. The temperature dependence of the coefficient $N_{eff}$ exhibits a maximum in the vicinity of the diffuse phase transition and approaches zero at increasing temperature (see Figure 6). Since the elastic stiffness does not possess any anomalies at the phase transition, one can expect that the anomalous dependence of $N_{eff}$ is associated only with a similar anomaly of the dielectric permitivity $\varepsilon_l$ (see, e.g., [12]). It is interesting to notice that the actual range for the coefficient $N^{(s)}_{eff}=N_{eff}\varepsilon_l(\varepsilon_l-1)=(1.12\sim1.31)\times10^{-9}$m$^4$/NC in the paraphase, re-calculated with the data shown in Figure 6, includes the value $(N^{(s)}_{421}+N^{(s)}_{431})=1.25\times10^{-9}$m$^4$/NC obtained on the basis of temperature dependence of the PO coefficient. Hence, one can see that our coefficients obtained in different experiments do coincide, in agreement with the predictions of phenomenology.

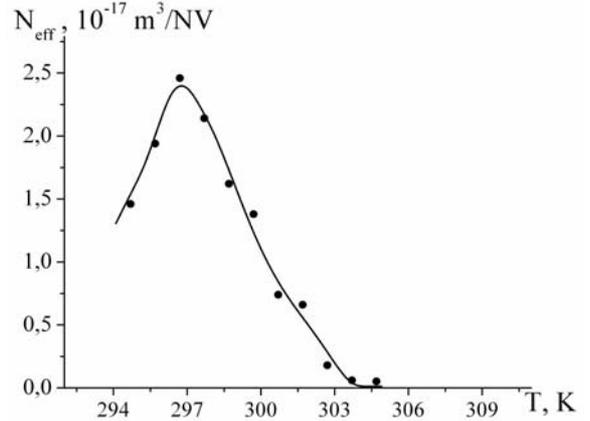

**Fig. 6.** Temperature dependence of the effective coefficient of combined PEO effect at the phase transition in Rochelle salt crystals ($\lambda$=632.8nm).

## Conclusion

In conclusion, our studies for the PEO effect in ferroelectric-ferroelastic Rochelle salt crystals have permitted us to calculate and compare the values of coefficients of the combined effect obtained from the measurements of the EO coefficient change under the action of mechanical stress and the PO coefficient change due to the effect of electric field (or spontaneous polarization). It has been experimentally shown that both coefficients are the same, as it is predicted by the theory. The temperature dependence of coefficient of the combined PEO effect is obtained. It has been also revealed that the change in the PO coefficients at the phase transition in Rochelle salt crystals may be satisfactorily explained by the linear and quadratic PEO effects induced by spontaneous polarization.

## Acknowledgement

The authors are grateful to the Ministry of Education and Science of Ukraine (the Project N0103U000701) for the financial support of this study.

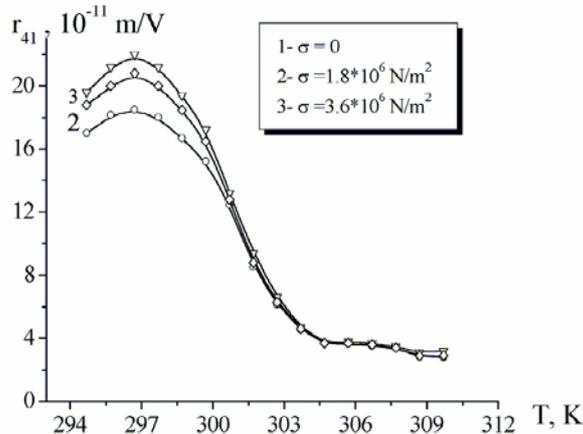

**Fig. 5.** Temperature dependence of the EO coefficient $r_{41}$ for NaKC$_4$H$_4$O$_6\times$4H$_2$NH$_2$O crystals at different mechanical stresses ($\lambda$=632.8nm).

## Errata

**Mys O., Adamiv V., Martynyuk-Lototska I. and Vlokh R. Ukr. J. Phys. Opt. 5 (2004) 6-7**.

Page 6. Last sentence of the first column was incorrectly printed. The correct sentence should read "They belong to the point group of symmetry $\bar{4}2m$."

**Martunyuk-Lototska I., Mys O., Dyachok Ya., Dudok T., Adamiv V., Burak Ya. and Vlokh R. Ukr.J.Phys.Opt. 5 (2004) 19-26.**

(i) Throughout the article, it should be $Li_2B_4O_7$ instead of $LiB_4O_7$.

(ii) Page 19. In the first sentence of the article it should be "…the point symmetry groups $\bar{3}m$ and 4mm." instead of "…the point symmetry groups 3m and 4mm.".

(iii) Page 24. The caption of Figure 2 should be: "Indicative surfaces of inverse acoustic wave velocities in the ABO (a) and LTB (b) crystals in the units [s/km]".

**Vlokh R. Ukr.J.Phys.Opt. 5 (2004) 27-31.**

Page 28. The correct form of Eqs. (3)–(8) is

$$n = 1 \pm 2\sqrt{\frac{GM}{c_0^4}}(\vec{g}_k^{1/2}) + \frac{3}{2}\left(\sqrt{\frac{GM}{c_0^4}}\right)^2 (\vec{g}_k^{1/2})(\vec{g}_l^{1/2})$$
$$\pm \frac{1}{2}\left(\sqrt{\frac{GM}{c_0^4}}\right)^3 (\vec{g}_k^{1/2})(\vec{g}_l^{1/2})(\vec{g}_m^{1/2}) + \frac{1}{16}\left(\sqrt{\frac{GM}{c_0^4}}\right)^4 (\vec{g}_k^{1/2})(\vec{g}_l^{1/2})(\vec{g}_m^{1/2})(\vec{g}_n^{1/2}) , \quad (3)$$

$$n = 1 \pm 2\sqrt{\frac{GM}{c_0^4}}(\vec{g}_k^{1/2}) + \frac{3}{2}\left(\sqrt{\frac{GM}{c_0^4}}\right)^2 (\vec{g}_k^{1/2})(\vec{g}_l^{1/2})$$
$$\pm \frac{1}{2}\left(\sqrt{\frac{GM}{c_0^4}}\right)^3 (\vec{g}_k^{1/2})(\vec{g}_l^{1/2})(\vec{g}_m^{1/2}) + \frac{1}{16}\left(\sqrt{\frac{GM}{c_0^4}}\right)^4 (\vec{g}_k^{1/2})(\vec{g}_l^{1/2})(\vec{g}_m^{1/2})(\vec{g}_n^{1/2}) , \quad (4)$$

$$n = 1 \pm 2\sqrt{\frac{GM}{c_0^4}}(\vec{g}_k^{1/2}) + \frac{3}{2}\left(\sqrt{\frac{GM}{c_0^4}}\right)^2 (\vec{g}_k^{1/2})(\vec{g}_l^{1/2}) , \quad (5)$$

$$B_{ij} = 1 \mp 4\sqrt{\frac{GM}{c_o^4}}(\vec{g}_k^{1/2}) - 7\left(\sqrt{\frac{GM}{c_0^4}}\right)^2 (\vec{g}_k^{1/2})(\vec{g}_l^{1/2}) , \quad (6)$$

$$n = 1 \pm 2\sqrt{\beta_{ij}M}(\vec{g}^{1/2}) + \frac{3}{2}\beta_{ijkl}M(\vec{g}_k^{1/2})(\vec{g}_l^{1/2}) , \quad (7)$$

$$B_{ij} = 1 \mp 4\sqrt{\beta_{ij}M}(\vec{g}^{1/2}) - 7\beta_{ijkl}M(\vec{g}_k^{1/2})(\vec{g}_l^{1/2}) . \quad (8)$$

Page 31. Last sentence of the paragraph "Refractive index dependence on the gravitation field" should be: "Therefore, due to the *Curie* principle, the symmetry group of the flat space should depend on the field configuration and, following the *Neumann* symmetry principle, it should be a subgroup of symmetry group of the time."